\documentclass[conference]{IEEEtran}
\IEEEoverridecommandlockouts
\usepackage{cite,amsmath,amssymb,amsfonts,algorithmic,graphicx,textcomp,xcolor,array,booktabs}
\graphicspath{{images/}}

\newcolumntype{C}[1]{>{\centering\let\newline\\\arraybackslash\hspace{0pt}}m{#1}}
\def\BibTeX{{\rm B\kern-.05em{\sc i\kern-.025em b}\kern-.08em
    T\kern-.1667em\lower.7ex\hbox{E}\kern-.125emX}}
    
\begin{document}

\title{Automatic 3D Ultrasound Segmentation of Uterus Using Deep Learning\\
\thanks{We acknowledge the support of the Natural Sciences and Engineering Research Council of Canada (NSERC)and Cancer Research UK  (A23557), and NHS funding to the NIHR Biomedical Research Centre at The Royal Marsden and The Institute of Cancer Research}}

\author{
\IEEEauthorblockN{1\textsuperscript{st} Bahareh Behboodi}
\IEEEauthorblockA{\textit{Department of Electrical and} \\
\textit{Computer Eng, Concordia University}\\
Montreal, Canada \\
b\_behboo@encs.concordia.ca}
\and
\IEEEauthorblockN{2\textsuperscript{nd} Hassan Rivaz}
\IEEEauthorblockA{\textit{Department of Electrical and} \\
\textit{Computer Eng, Concordia University}\\
Montreal, Canada \\
hrivaz@ece.concordia.ca}
\and
\IEEEauthorblockN{3\textsuperscript{rd} Susan Lalondrelle}
\IEEEauthorblockA{\textit{Institute of Cancer Research} \\
London, United Kingdom \\
susan.lalondrelle@rmh.nhs.uk}
\and
\IEEEauthorblockN{4\textsuperscript{th} Emma Harris}
\IEEEauthorblockA{\textit{Institute of Cancer Research}\\
London, United Kingdom \\
Emma.Harris@icr.ac.uk}
}

\maketitle

\begin{abstract}
On-line segmentation of the uterus can aid effective image-based guidance for precise delivery of dose to the target tissue (the uterocervix) during cervix cancer radiotherapy. 3D ultrasound (US) can be used to image the uterus, however, finding the position of uterine boundary in US images is a challenging task due to large daily positional and shape changes in the uterus, large variation in bladder filling, and the limitations of 3D US images such as low resolution in the elevational direction and imaging aberrations. Previous studies on uterus segmentation mainly focused on developing semi-automatic algorithms where require manual initialization to be done by an expert clinician. Due to limited studies on the automatic 3D uterus segmentation, the aim of the current study was to overcome the need for manual initialization in the semi-automatic algorithms using the recent deep learning-based algorithms. Therefore, we developed 2D UNet-based networks that are trained based on two scenarios. In the first scenario, we trained 3 different networks on each plane (i.e., sagittal, coronal, axial) individually. In the second scenario, our proposed network was trained using all the planes of each 3D volume. Our proposed schematic can overcome the initial manual selection of previous semi-automatic algorithm.
\end{abstract}

\begin{IEEEkeywords}
Uterus segmentation, Deep learning, Ultrasound
\end{IEEEkeywords}

\section{Introduction}
Cervical cancer as one of the most frequent cancer types in women, affects more than half a million females each year and results in 300 000 deaths world wide~\cite{cohen2019cervical}. It is, however, largely preventable, and the treatment is dependent on the severity of the condition and availability of local resources at the time of diagnosis~\cite{cohen2019cervical}. Recent studies have shown that incorporating the results of advanced imaging technology and surgical staging lead to more enhanced prognosis and treatment planning~\cite{hsiao2021updated}. Imaging modalities such as magnetic resonance imaging (MRI), computed tomography (CT), positron emission tomography (PET), and ultrasound (US) imaging have been utilised for treatment plans. However, MRI, CT, and PET imaging facilities, are costly, not uniformly available, require a long scanning time, and are not real-time. Thus, US imaging has emerged as the most suited modality for cervical cancer screening due to its cost-effectiveness, radiation-free, non-invasiveness, ease of use on the bedside, and real-time nature.

\begin{figure}[h]
  \centering
  \centerline{\includegraphics[width=9cm]{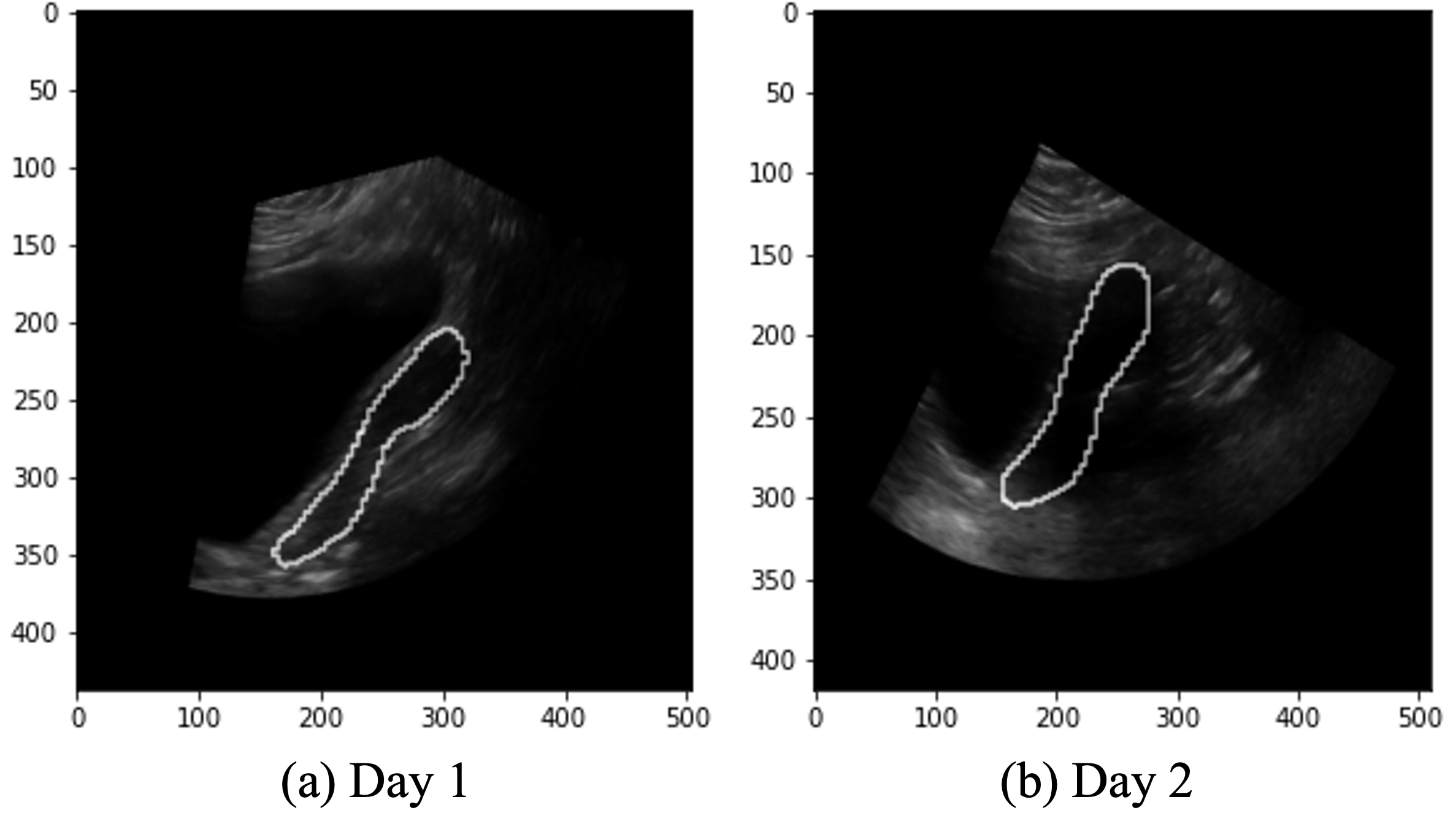}}
  \caption{An illustration of uterine location variation (sagittal view) in one patient across two scans taken on different days.}
  \label{fig:vari}
\end{figure}

Radiotherapy is a type of treatment that delivers a dose of radiation to the target tissues, however, its effect and efficiency in the treatment of cervical cancer is limited by motion of the target tissues~\cite{huh2004interfractional}. Therefore, on-line segmentation of the uterus can aid effective image-based guidance for precise delivery of dose to the target tissue (the uterocervix) during cervix cancer radiotherapy. Furthermore, segmenting the uterus can aid in determining the extent of a tumour and the presence of metastatic disease. However, finding the position of uterine boundary in US images is a challenging task due to large daily positional and shape changes in the uterus (shown in Fig. \ref{fig:vari}), large variation in bladder filling, and the limitations of 3D US images such as low resolution in the elevational direction. One group of studies on uterus segmentation mainly focused on developing semi-automatic algorithms where require manual initialization to be done by an expert clinician. Mason \textit{et al.} developed a semi-automatic algorithm such that a central sagittal plane is manually contoured. Then, the selected plane and contour are used as a start point of fitting elliptical contours in semi-axial planes~\cite{mason2020stacked}. Another group, focused on use of conventional image processing techniques for automatic detection and segmentation of uterine fibroid~\cite{dilna2020fibroid,padghamod2014classification}. 

Recent advances in image processing approaches, such as artificial intelligence (AI) and deep learning (DL) algorithms, have paved the way in solving a variety of problems. AI and deep learning approaches in medicine have a lot of potential, particularly in US diagnostic imaging, where large datasets must be managed. In US image analysis, many researchers have shown the promising results in detection of breast lesions~\cite{behboodi2021deep,amiri2020two,behboodi2019ultrasound,tehrani2021ultrasound}, muscle ~\cite{loram2020objective}, thyroid nodule ~\cite{ouahabi2021deep}, prostate\cite{shi2016stacked}, liver~\cite{wu2014deep}, brain~\cite{sombune2017automated}. However, due to limited studies on the automatic segmentation of uterus US images, the main focus of the current study is to investigate more on automatic segmentation of 3D uterus US images and to eliminate the need for manual initialization in the previous semi-automatic algorithms using the recent deep learning-based techniques. Deep learning techniques' success is heavily dependent on the amount of available data with annotations, and creating annotations for US pictures is a time and money-intensive operation. To be more explicit, 3D networks have higher number of parameters, which causes memory issues and a greater demand for annotated 3D data. Therefore, due to limited available 3D uterus data, we explore 2D networks that use 2D planes of 3D volumes.

\section{Materials and Methods}
\subsection{Dataset}
\begin{figure}[h!]
  \centering
  \centerline{\includegraphics[width=9cm]{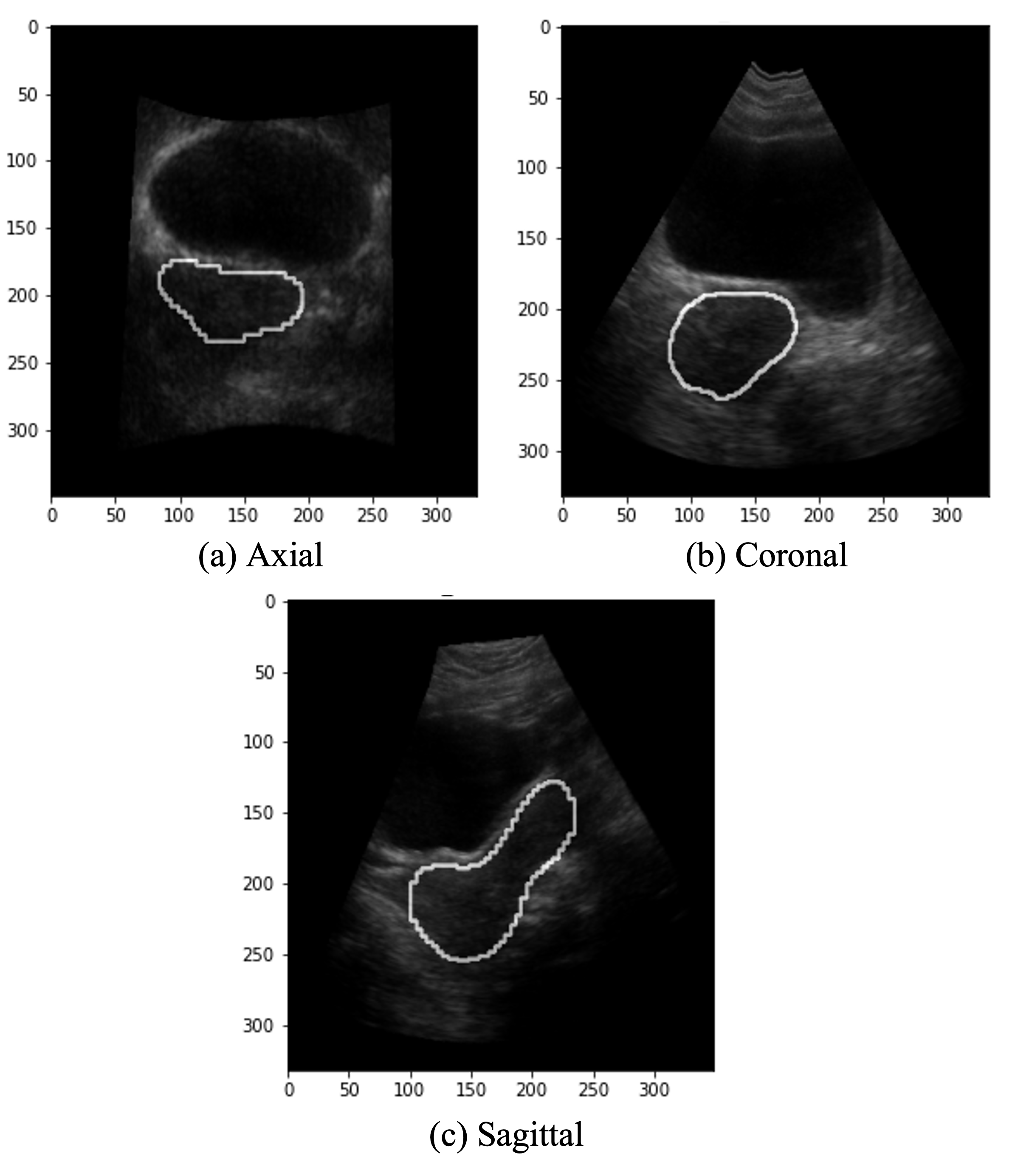}}
  \caption{An example of 3D US image with the uterus annotation across (a) axial, (b) coronal, and (c) sagittal planes.}
  \label{fig:img_mask}
\end{figure}

The dataset that used in the current study consist of 3D US images of 11 patients. On average, each patient received 4 sessions of 3D US scanning leading to a total of 38 3D US scans, with each 3D scan comprising $>$100 2D images. Two patients were chosen as the test set and the remainder as the train set, resulting in a total of 35 and 3 scans for the train and test sets, respectively. Table \ref{tab:data} presents the details of number of scans for each patient. An example of US images with their overlaid annotations across all planes (i.e. axial, coronal, sagittal) is presented in Fig. \ref{fig:img_mask}. We scaled all the scans to an identical shape 576$\times$576$\times$576 as the 3D volumes varied in size.
\begin{table}[h]
	\centering
	\caption{Number of 3D US scans per patient. Patient 1 and 10 were selected as the test set, and the rest were grouped as the train set.}
	\label{tab:data}
	\small
	\begin{tabular}{ C{2cm} C{2cm} C{2cm}}
		\toprule[0.5mm]
		\bf Patient ID & \bf No. 3D scans & \bf Train/Test\\
		\toprule
		1 & 2 & Test\\
		2 & 5 & Train\\
		3 & 3 & Train\\
		4 & 4 & Train\\
		5 & 5 & Train\\
		6 & 4 & Train\\
		7 & 5 & Train\\ 
		8 & 3 & Train\\ 
		9 & 5 & Train\\ 
		10 & 1 & Test\\
		11 & 1 & Train\\
		
		\toprule
	\end{tabular}
\end{table}

\subsection{Protocol}
Most of the recently developed deep learning algorithms suffer from generalization, and the performance of such algorithms for a new dataset need to be investigated. Furthermore, training 3D networks with only 38 3D volumes is not possible. Therefore, we developed 2D networks for segmentation and stacks the outputs into a 3D volume as the final prediction. Each 3D volume partitioned into 2D slices known as the coronal, sagittal, and axial planes. We proceeded our analysis through two main scenarios. In the first scenario, we trained 3 different 2D networks on each 2D plane (i.e., sagittal, coronal, axial) individually. In the second scenario, our proposed 2D network was trained using 2D images across all the planes of each 3D volume. 
\subsection{Experiments}
The proposed network was based on well-known segmentation architecture, U-Net~\cite{unet}, where its feature extractor is set to MobileNet-v2~\cite{howard2018inverted}. Segmentation masks generated using the proposed algorithm were compared to expert manual contours. We had three and one networks to train in the first and second scenarios, respectively. For simplicity, we refer to \textit{net\_X}, \textit{net\_Y}, \textit{net\_Z}, and \textit{net\_all} as networks trained on 2D images of axial, coronal, sagittal, and all planes. All the aforementioned networks trained for 200 epochs, using Adam~\cite{adam} optimizer with learning rate $1e-4$ and weight decay $0.025$. 2D images were reshaped to the size of 576$\times$576 where a center crop augmentation with the cropping window size of 512$\times$512 were applied as the augmentation. Additionally, images were flipped vertically and/or horizontally on random basis. 5-fold cross validation was conducted to prevent variation in networks performance. The loss function was set to the combination of binary cross-entropy (BCE) and dice similarity (DSC) functions (Eq. \ref{eq:loss}).
\begin{equation}
loss = BCE + 0.5 * DSC
\label{eq:loss}
\end{equation}
where $DSC = \frac{2 |G\cap P| + ep}{|G|+|P|+ep}$, $G$ and $P$ is ground truth and predicted segmentation masks, respectively, and $ep=1$. And, $BCE = ylog(P(y)) + (1-y)log(1-P(y)$, where $y$ and $P$ denote predictions and probability function, respectively.

%
\section{Results}
\begin{figure}[h!]
  \centering
  \centerline{\includegraphics[width=9cm]{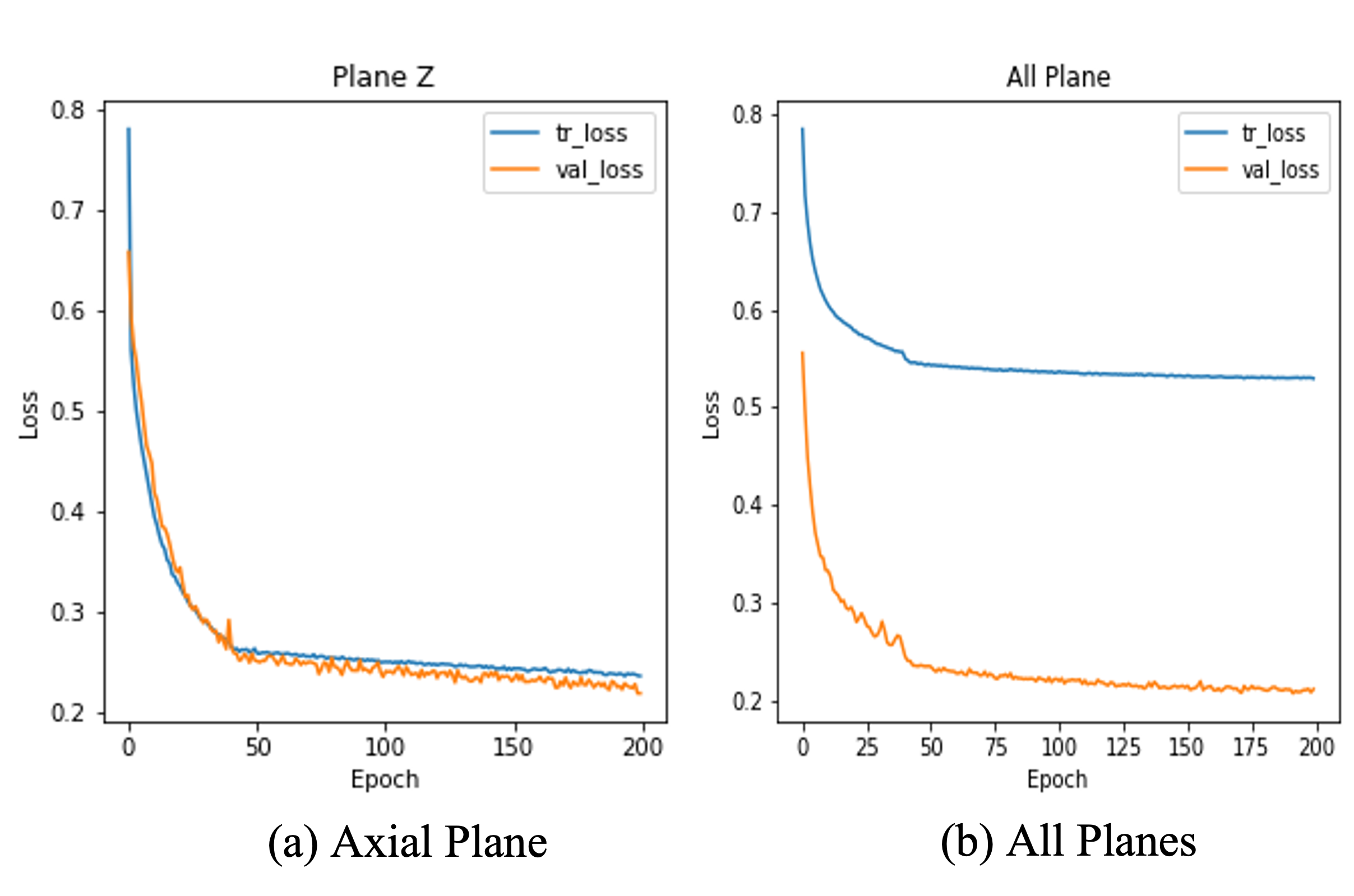}}
  \caption{Train\_validation loss for the 1st fold of 5-fold cross validation. ((a)-(c): 1st scenario, (d): 2nd scenario).}
  \label{fig:tr_val}
\end{figure}

\begin{figure}[h!]
  \centering
  \centerline{\includegraphics[width=8.5cm]{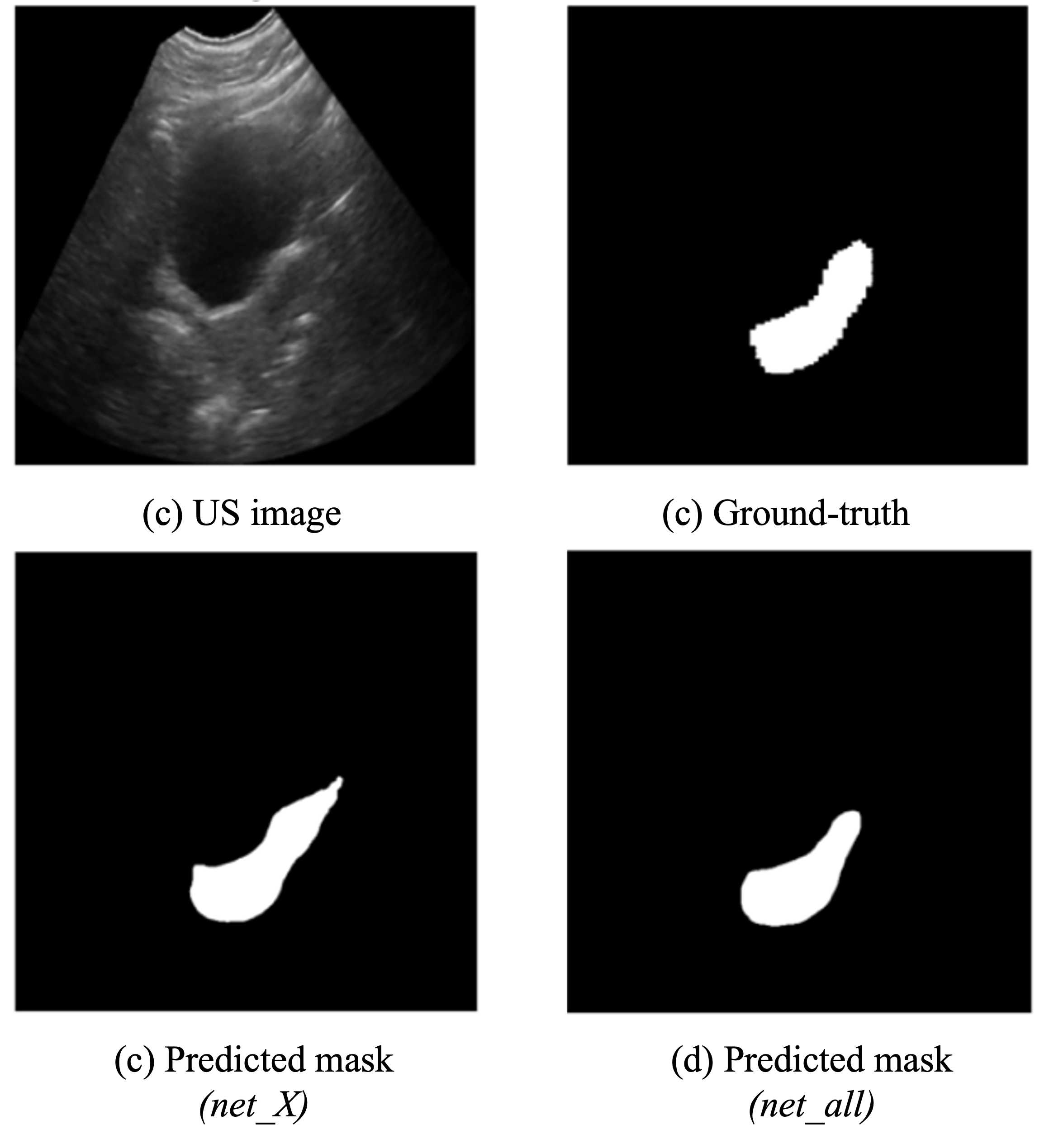}}
  \caption{An example of ground truth versus predicted segmentation masks from \textit{net\_X} (DSC=0.88) and \textit{net\_all} (DSC=0.8) for a middle slice.}
  \label{fig:res}
\end{figure}

Figure \ref{fig:tr_val} shows the train-validation loss across 2 networks. We only include the train\_validation loss of \textit{net\_X} due to similarity of train\_validation loss in other networks (i.e. \textit{net\_Y} and \textit{net\_Z}) of our 1st scenario. We observed that when we combine all the planes of 3D volume (axial, coronal, and sagittal), Fig. \ref{fig:tr_val} (b). Figure \ref{fig:res} (c), and (d) show an example of sagittal slice of one patient, where the uterus is fully visible, predicted from \textit{net\_X} and \textit{net\_all} based on our first and second scenarios, respectively. 

\begin{figure}[h!]
  \centering
  \centerline{\includegraphics[width=9cm]{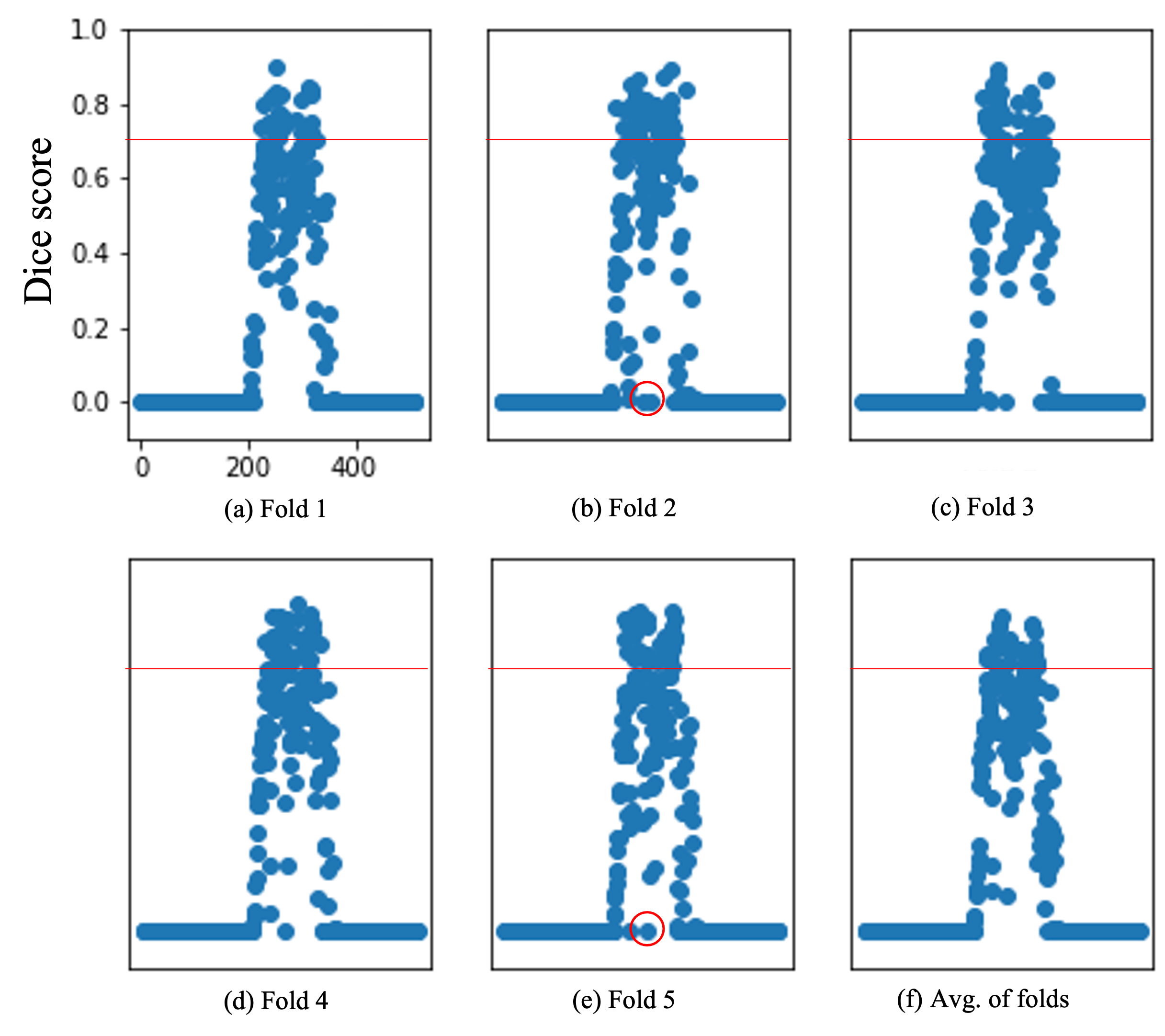}}
  \caption{Distribution of the DSC across folds for patient ID 1.}
  \label{fig:dsc}
\end{figure}

We observed that for the middle slices where the uterus is fully visible, the DSC is high for both test patients. However, for the slices close to the edges of uterus, the DSC is low that means the network performs well mainly on middle slices. The distribution of the DSC across slices in the axial plane for one scan in all 5 folds is illustrated in Fig. \ref{fig:dsc}. The distribution of DSC in each fold is shown in (a)-(e), and the average of DSC is shown in (f). The red line in this figure shows the DSC of 0.7. Therefore, our proposed algorithms can overcome the need of manual selection of the middle slices for the semi-automatic presented in Mason \textit{et al.}\cite{mason2020stacked}. Some slices, however, are in the middle and have a low DSC (marked with red circles in Fig. \ref{fig:dsc}). In the future, we will look at these cases more.

The quantitative results are reported in Table \ref{tab:res}. The average DSC for most scans is low due to the difficulty in segmenting slices on the edges that we addressed earlier. However, we observed that the DSC of middle slices are higher as we expected, and both scenarios behave pretty similarly.

\begin{table}[h]
	\centering
	\caption{Quantitative results - Average DSC - Scenario 1.}
	\label{tab:res}
	\small
	\begin{tabular}{ C{1cm} C{0.5cm} C{1.5cm} C{1.5cm} C{1.5cm}}
		\toprule[0.5mm]
		\bf Patient ID & \bf Scan No. & \bf \textit{net\_X} & \bf \textit{net\_Y} & \bf \textit{net\_Z}\\
		\toprule
		\multicolumn{5}{c}{All slices}\\
		\toprule
		1 & 1 &$0.48\pm0.24$ & $0.61\pm0.19$&$0.37\pm0.15$\\
		  & 2 &$0.55\pm0.21$ & $0.45\pm0.2$ &$0.44\pm0.17$\\
		\midrule
		10 & 1 &$0.58\pm0.21$&$0.69\pm0.24$&$0.58\pm0.24$\\
		
		\toprule
		\multicolumn{5}{c}{4 mid-slices}\\
		\toprule
		1 & 1 &$0.68\pm0.1$ &$0.72\pm0.05$&$0.56\pm0.08$\\
		  & 2 &$0.64\pm0.05$ &$0.29\pm0.04$&$0.46\pm0.08$\\
		\midrule
		10 & 1 &$0.67\pm0.11$ &$0.85\pm0.03$&$0.67\pm0.11$\\
		\toprule
	\end{tabular}
\end{table}
\begin{table}[h]
	\centering
	\caption{Quantitative results - Average DSC - Scenario 2.}
	\label{tab:res}
	\small
	\begin{tabular}{ C{1cm} C{0.5cm} C{1.5cm} C{1.5cm} C{1.5cm}}
		\toprule[0.5mm]
		\bf Patient ID & \bf Scan No. & \bf Axial & \bf Coronal & \bf Sagittal\\
		\toprule
		\multicolumn{5}{c}{All slices}\\
		\toprule
		1 & 1 &$0.55\pm0.27$ & $0.61\pm0.21$&$0.37\pm0.16$\\
		  & 2 &$0.53\pm0.20$ & $0.42\pm0.21$ &$0.44\pm0.21$\\
		\midrule
		10 & 1 &$0.56\pm0.22$&$0.64\pm0.32$&$0.60\pm0.25$\\
		
		\toprule
		\multicolumn{5}{c}{4 mid-slices}\\
		\toprule
		1 & 1 &$0.77\pm0.05$ &$0.59\pm0.11$&$0.62\pm0.08$\\
		  & 2 &$0.64\pm0.06$ &$0.32\pm0.06$&$0.45\pm0.06$\\
		\midrule
		10 & 1 &$0.63\pm0.11$ &$0.84\pm0.02$&$0.71\pm0.03$\\
		\toprule
	\end{tabular}
\end{table}
\section{Discussion}
As mentioned earlier, uterus segmentation in US images is very challenging due to its location and inconspicuous boundaries.  In the previous semi-automatic algorithm presented by Mason \textit{et al.} \cite{mason2020stacked}, the start point of the algorithm is finding the slice where the uterus is completely visible. Therefore, our proposed schematic not only overcome the initial manual selection of previous semi-automatic algorithm, it also provides comparable DSC with the semi-automatic algorithm. As we utilized MobileNet-v2 which is well-known in terms of being light in memory usage, the proposed network configuration is also sufficiently light which makes it suitable for use in the clinic which requires results in a few seconds. We discovered that all of the proposed networks function inadequately on slices close to the uterus's boundaries, which is a shortcoming of the current study. As part of our ongoing research, we will delve deeper into this issue.

\bibliographystyle{IEEEbib}
\bibliography{refs}

\end{document}